\documentclass[journal]{IEEEtran}
\usepackage[T1]{fontenc}
\usepackage{subfigure, graphicx, caption, multirow,stmaryrd, dsfont}
\usepackage{graphicx, booktabs, subfigure, multirow, amssymb, amsmath, amsthm, algorithm, algorithmic, threeparttable}
\usepackage{xcolor, cite}
\usepackage[normalem]{ulem}
\usepackage[makeroom]{cancel}
\usepackage{threeparttable}
\ifCLASSINFOpdf
\else
\fi
\usepackage{amsmath}
\usepackage[cmintegrals]{newtxmath}
\begin{document}
%
\title{CFR-RL: Traffic Engineering with Reinforcement Learning in SDN}
%
%
%
\author{Junjie~Zhang,~\IEEEmembership{Member,~IEEE},~Minghao~Ye,~Zehua~Guo,~\IEEEmembership{Senior Member,~IEEE},

~Chen-Yu~Yen,~and~H.~Jonathan~Chao,~\IEEEmembership{Fellow,~IEEE}
\thanks{The work of Z. Guo was supported in part by National Key Research and Development Program of China under Grant 2018YFB1003700 and Beijing Institute of Technology Research Fund Program for Young Scholars. (Corresponding author: Zehua Guo)}
\thanks{J. Zhang is with Fortinet, Inc., Sunnyvale, CA 94086 USA (e-mail: junjie.zhang@nyu.edu).}
\thanks{M. Ye, C.-Y. Yen and H. J. Chao are with the Department of Electrical and Computer
Engineering, New York University, New York City, NY 11201 USA (e-mail: my1706@nyu.edu; cyy310@nyu.edu; chao@nyu.edu).}
\thanks{Z. Guo is with Beijing Institute of Technology, Beijing 100081, China (e-mail: guo@bit.edu.cn).}}

\maketitle

\begin{abstract}
Traditional Traffic Engineering (TE) solutions can achieve the optimal or near-optimal performance by rerouting as many flows as possible. However, they do not usually consider the negative impact, such as packet out of order, when frequently rerouting flows in the network. To mitigate the impact of network disturbance, one promising TE solution is forwarding the majority of traffic flows using Equal-Cost Multi-Path (ECMP) and selectively rerouting a few \textit{critical flows} using Software-Defined Networking (SDN) to balance link utilization of the network. However, critical flow rerouting is not trivial because the solution space for critical flow selection is enormous. Moreover, it is impossible to design a heuristic algorithm for this problem based on fixed and simple rules, since rule-based heuristics are unable to adapt to the changes of the traffic matrix and network dynamics. In this paper, we propose CFR-RL (Critical Flow Rerouting-Reinforcement Learning), a Reinforcement Learning-based scheme that learns a policy to select critical flows for each given traffic matrix automatically. CFR-RL then reroutes these selected critical flows to balance link utilization of the network by formulating and solving a simple Linear Programming (LP) problem. Extensive evaluations show that CFR-RL achieves near-optimal performance by rerouting only 10\%-21.3\% of total traffic.
\end{abstract}

\begin{IEEEkeywords}
Reinforcement Learning, Software-Defined Networking, Traffic Engineering, Load Balancing, Network Disturbance Mitigation.
\end{IEEEkeywords}

%
\IEEEpeerreviewmaketitle

\section{Introduction} %
\label{intro}

The emerging Software-Defined Networking (SDN) provides new opportunities to improve network performance \cite{mckeown2008openflow}. In SDN, the control plane can generate routing policies based on its global view of the network and deploy these policies in the network by installing and updating flow entries at the SDN switches.

Traffic Engineering (TE) is one of important network features for SDN \cite{agarwal2013traffic, guo2014traffic, zhang2014dynamic}, and is usually implemented in the control plane of SDN. The goal of TE is to help Internet Service Providers (ISPs) optimize network performance and resource utilization by configuring the routing across their backbone networks to control traffic distribution \cite{zhang2015load, guo2019joint}. Due to dynamic load fluctuation among the nodes, traditional TE \cite{wang1999explicit, osborne2002traffic, fortz2002optimizing, holmberg2004optimization, chu2009optimal, zhang2012optimizing} reroutes many flows periodically to balance the load on each link to minimize network congestion probability, where a flow is defined as a source-destination pair. One usually formulates the flow routing problem with a particular performance metric as a specific objective function for optimization. For a given traffic matrix, one often wants to route all the flows in such a way that the maximum link utilization in the network is minimized. 

Although traditional TE solutions can achieve the optimal or near-optimal performance by rerouting as many flows as possible, they do not consider the negative impact, such as packet out of order, when rerouting the flows in the network. To reach the optimal performance, TE solutions might reroute many traffic flows to just slightly reduce the link utilization on the most congested link, leading to significant network disturbance and service disruption. For example, a flow between two nodes in a backbone network is aggregated of many micro-flows (e.g., five tuples-based TCP flows) of different applications. Changing the path of a flow could temporarily affect many TCP flows' normal operation. Packets loss or out-of-order may cause duplicated ACK transmissions, triggering the sender to react and reduce its congestion window size and hence decrease its sending rate, eventually increasing the flow's completion time and degrading the flow's Quality of Service (QoS). In addtion, rerouting all flows in the network could incur a high burden on the SDN controller to calculate and deploy new flow paths \cite{zhang2014dynamic}. Because rerouting flows to reduce congestion in backbone networks could adversely affect the quality of users' experience, network operators have no desire to deploy these traditional TE solutions in their networks unless reducing network disturbance is taken into the consideration in designing the TE solutions.

To mitigate the impact of network disturbance, one promising TE solution is forwarding majority of traffic flows using Equal-Cost Multi-Path (ECMP) and selectively rerouting a few \textit{critical flows} using SDN to balance link utilization of the network, where a critical flow is defined as a flow with a dominant impact to network performance  (e.g., a flow on the most congested link) \cite{zhang2014dynamic, zhang2014hybrid_routing}. Existing works show that critical flows exist in a given traffic matrix \cite{zhang2014dynamic}. ECMP reduces the congestion probability by equally splitting traffic on equal-cost paths while critical flow rerouting aims to achieve further performance improvement with low network disturbance. 

The critical flow rerouting problem can be decoupled into two sub-problems: (1) identifying critical flows and (2) rerouting them to achieve good performance. Although sub-problem (2) is relatively easy to solve by formulating it as a Linear Programming (LP) optimization problem, solving sub-problem (1) is not trivial because the solution space is huge. For example, if we want to find 10 critical flows among 100 flows, the solution space has $C_{100}^{10} \approx 17$ trillion combinations. Considering the fact that traffic matrix varies in the level of minutes, an efficient solution should be able to quickly and effectively identify the critical flows for each traffic matrix. Unfortunately, it is impossible to design a heuristic algorithm for the above algorithmically-hard problem based on fixed and simple rules. This is because rule-based heuristics are unable to adapt to the changes of the traffic matrix and network dynamics and thus unable to guarantee their performance when their design assumptions are violated, as later shown in Section \ref{sec:evaluation}.

In this paper, we propose CFR-RL (Critical Flow Rerouting-Reinforcement Learning), a Reinforcement Learning-based scheme that performs critical flow selection followed by rerouting with linear programming. CFR-RL learns a policy to select critical flows purely through observations, without any domain-specific rule-based heuristic. It starts from scratch without any prior knowledge, and gradually learns to make better selections through reinforcement, in the form of reward signals that reflects network performance for past selections. By continuing to observe the actual performance of past selections, CFR-RL would optimize its selection policy for various traffic matrices as time goes. Once training is done, CFR-RL will efficiently and effectively select a small set of critical flows for each given traffic matrix, and reroute them to balance link utilization of the network by formulating and solving a simple linear programming optimization problem. 

The main contributions of this paper are summarized as follows:
\begin{enumerate}
\item We consider the impact of flow rerouting to network disturbance in our TE design and propose an effective scheme that not only minimizes the maximum link utilization but also reroutes only a small number of flows to reduce network disturbance. 
\item We customize a RL approach to learn the critical flow selection policy, and utilize LP as a reward function to generate reward signals. This RL$+$LP combined approach turns out to be surprisingly powerful.
\item We evaluate and compare CFR-RL with other rule-based heuristic schemes by conducting extensive experiments on different topologies with both real and synthesized traffic. CFR-RL not only outperforms rule-based heuristic schemes by up to 12.2\%, but also reroutes 11.4\%-14.7\% less traffic on average. Overall, CFR-RL is able to achieve near-optimal performance by rerouting only 10\%-21.3\% of total traffic. In addition, the evalution results show that CFR-RL is able to generalize to unseen traffic matrices.
\end{enumerate}

The remainder of this paper is organized as follows. Section II describes the related works. Section III presents the system design. Section IV discusses how to train the critical flow selection policy using a RL-based approach. Section V describes how to reroute the critical flows. Section VI evaluates the effectiveness of our scheme. Section VII concludes the paper and discusses future work.

\section{Related Works} \label{relatedworks}

\subsection{Traditional TE Solutions}
In Multiprotocol Label Switching (MPLS) networks, a routing problem has been formulated as an optimization problem where explicit routes are obtained for each source-destination pair to distribute traffic flows \cite{wang1999explicit, osborne2002traffic}. Using Open Shortest Path First (OSPF) and ECMP protocols, \cite{fortz2002optimizing, holmberg2004optimization, chu2009optimal} attempt to balance link utilization as even as possible by carefully tuning the link costs to adjust path selection in ECMP. OSPF-OMP (OMP, Optimized Multipath) \cite{OSPF-OMP}, a variation of OSPF, attempts to dynamically determine the optimal allocation of traffic among multiple equal-cost paths based on the exchange of special traffic-load control messages. Weighted ECMP \cite {zhang2012optimizing} extends ECMP to allow weighted traffic splitting at each node and achieves significant performance improvement over ECMP. Two-phase routing optimizes routing performance by selecting a set of intermediate nodes and tuning the traffic split ratios to the nodes \cite{kodialam2008oblivious, antic2009two}. In the first phase, each source sends traffic to the intermediate nodes based on predetermined split ratios, and in the second phase, the intermediate nodes then deliver the traffic to the final destinations. This approach requires IP tunnels, optical-layer circuits, or label switched paths in each phase.

\subsection{SDN-Based TE Solutions}
Thanks to the flexible routing policy from the emerging SDN, dynamic hybrid routing \cite{zhang2014dynamic} achieves load balancing for a wide range of traffic scenarios by dynamically rebalancing traffic to react to traffic fluctuations with a preconfigured routing policy. Agarwal et al. \cite{agarwal2013traffic} consider a network with partially deployed SDN switches. They improve network utilization and reduce packet loss by strategically placing the controller and SDN switches. Guo et al. \cite{guo2014traffic} propose a novel algorithm named SOTE to minimize the maximum link utilization in an SDN/OSPF hybrid network.

\subsection{Machine Learning-Based TE Solutions}
Machine learning has been used to improve the performance of backbone networks and data center networks. For backbone networks, Geyer et al. \cite{geyer2018learning} design an automatic network protocol using semi-supervised deep learning. Sun et al. \cite{sun2019sinet} selectively control a set of nodes and use a RL-based policy to dynamically change the routing decision of flows traversing the selected nodes. To minimize signaling delay in large SDNs, Lin et al. \cite{lin2016qos} employ a distributed three-level control plane architecture coupled with a RL-based solution named QoS-aware Adaptive Routing. Xu et al.  \cite{xu2018experience} use RL to optimize the throughput and delay in TE. AuTO \cite{chen2018auto} is developed to optimize routing traffic in data center networks with a two-layer RL. One is called the Peripheral System for deploying hosts and routing small flows, and the other one is called the Central System for collecting global traffic information and routing large flows. 

However, all of the above works do not consider mitigating the impact of network disturbance and service disruption caused by rerouting. 

\section{System Design}

In this section, we describe the design of CFR-RL, a RL-based scheme that learns a critical flow selection policy and reroutes the corresponding critical flows to balance link utilization of the network. 

\begin{figure}[t]
\centering
\includegraphics[width=0.3\textwidth]{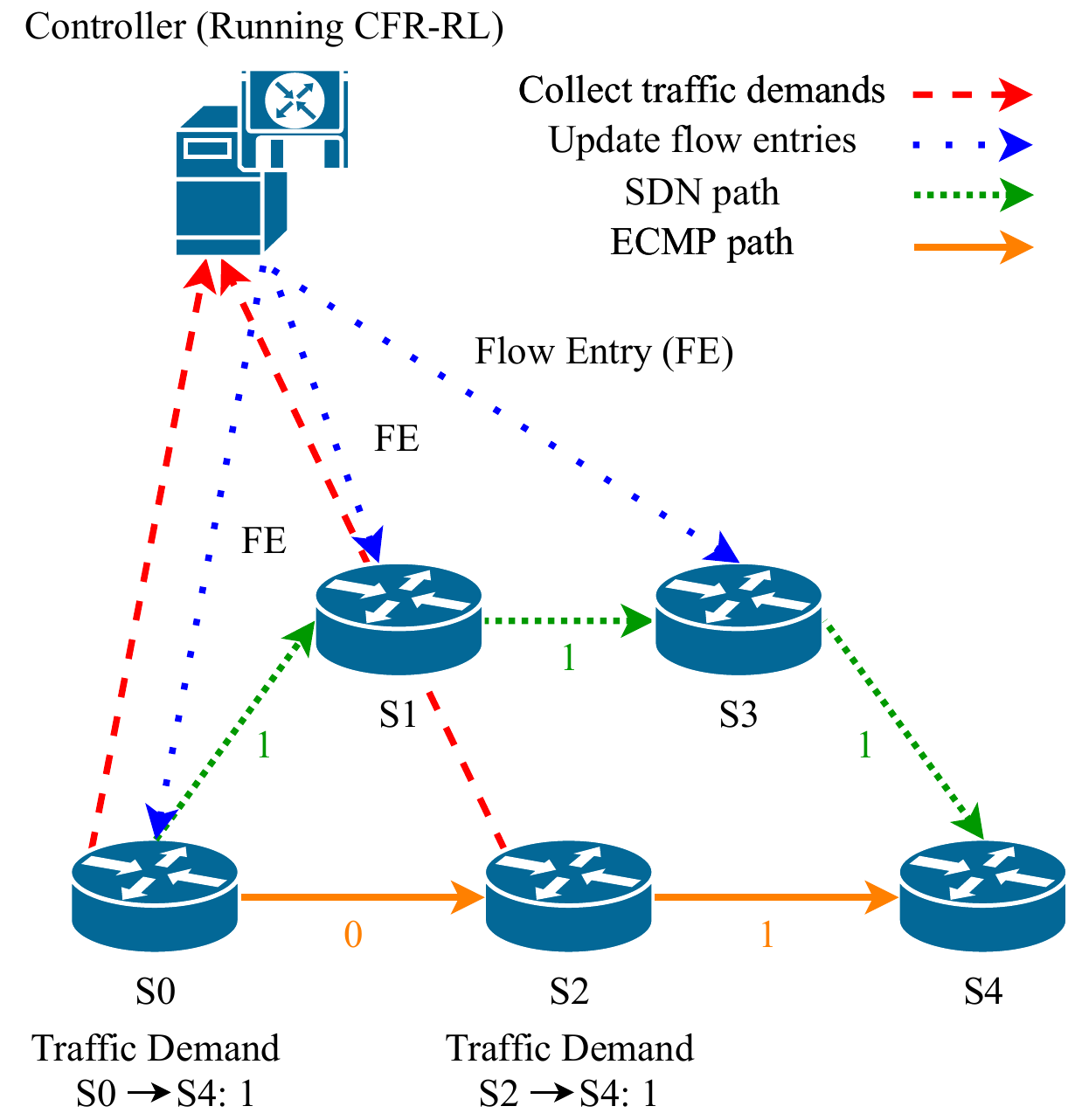}
\caption{An illustrative example of CFR-RL rerouting procedure. Each link capability equal to 1. Best viewed in color.}
\label{fig:system}
\end{figure}

We train CFR-RL to learn a selection policy over a rich variety of historical traffic matrices, where traffic matrices can be measured by SDN switches and collected by an SDN central controller periodically \cite{xu2017minimizing}. CFR-RL represents the selection policy as a neural network that maps a "raw" observation (e.g., a traffic matrix) to a combination of critical flows. The neural network provides a scalable and expressive way to incorporate various traffic matrices into the selection policy. CFR-RL trains this neural network based on REINFORCE algorithm \cite{Williams1992} with some customizations, as detailed in Section \ref{sec: learning}. 

Once training is done, CFR-RL applies the critical flow selection policy to each real time traffic matrix provided by the SDN controller periodically, where a small number of critical flows (e.g., $K$) are selected. The evaluation results in Section \ref{sec:critical_flows_num} show that selecting 10\% of total flows as critical flows (roughly 11\%-21\% of total traffic) is sufficient for CFR-RL to achieve near-optimal performance, while network disturbance (i.e., the percentage of total rerouted traffic) is reduced by at least $78.7\%$ compared to rerouting all flows by traditional TE. Then the SDN controller reroutes the selected critical flows by installing and updating corresponding flow entries at the switches using a flow rerouting optimization method described in Section \ref{sec:rerouting}. The remaining flows would continue to be routed by the default ECMP routing. Note that the flow entries at the switches for the critical flows selected in the previous period will time out, and the flows would be routed by either default ECMP routing or newly installed flow entries in the current period. Figure \ref{fig:system} shows an illustrative example. CFR-RL reroutes the flow from S0 to S4 to balance link load by installing forwarding entries at the corresponding switches along the SDN path.

There are two reasons we do not want to adopt RL for the flow rerouting problem. Firstly, since the set of critical flows is small, LP is an efficient and optimal method to solve the rerouting problem. Secondly, a routing solution consists of a split ratio (i.e., traffic demand percentage) for each flow on each link. Given a network with $E$ links, there will be total $E*K$ split ratios in the routing solution, where $K$ is the number of critical flows. Since split ratios are continuous numbers, we have to adopt the RL methods for continuous action domain \cite{ddpg, John2015TRPO}. However, due to the high-dimensional, continuous action spaces, it has been shown that this type of RL methods would lead to slow and ineffective learning when the number of output parameters (i.e., $E*K$) is large \cite{xu2018experience, Valadarsky2017learning}.

\section{Learning A Critical Flow Selection Policy} \label{sec: learning}

In this section, we describe how to learn a critical flow selection policy using a customized RL approach. 

\subsection{Reinforcement Learning Formulation}
\medbreak
\noindent {\bf Input / State Space:} An agent takes a state $s_t = TM_t$ as an input, where $TM_t$ is a traffic matrix at time step $t$ that contains information of traffic demand of each flow. Typically, the network topology remains unchanged. Thus, we do not include the topology information as a part of the input. The results in Section \ref{sec:evaluation} show that CFR-RL is able to learn a good policy $\pi$ without prior knowledge of the network. It is worth noting that including additional information like link states as a part of input might be beneficial for training the critical flow selection policy. We will investigate it in our future work.

\medbreak
\noindent {\bf Action Space:} For each state $s_t$, CFR-RL would select $K$ critical flows. Given that there are total $N*(N-1)$ flows in a network with $N$ nodes, this RL problem would require a large action space of size $C^K_{N*(N-1)}$. Inspired by \cite{Mao:2016:RMD:3005745.3005750}, we define the action space as \{0, 1, ..., $(N*(N-1))-1$\} and allow the agent to sample $K$ different actions in each time step $t$ (i.e., $a_t^1, a_t^2, ..., a_t^K$).

\medbreak
\noindent {\bf Reward:} After sampling $K$ different critical flows (i.e., $f_K$) for a given state $s_t$, CFR-RL reroutes these critical flows and obtains the maximum link utilization $U$ by solving the rerouting optimization problem (\ref{eq:objective}) (described in the following section). Reward $r$ is defined as $1/U$, which is set to reflect the network performance after rerouting critical flows to balance link utilization. The smaller $U$ (i.e., the greater reward $r$), the better performance. In other words, CFR-RL adopts LP as a reward function to produce reward signals $r$ for RL.

\subsection{Training Algorithm} \label{sec:training_algorithm}

\begin{figure}[h]
\centering
\includegraphics[width=0.38\textwidth]{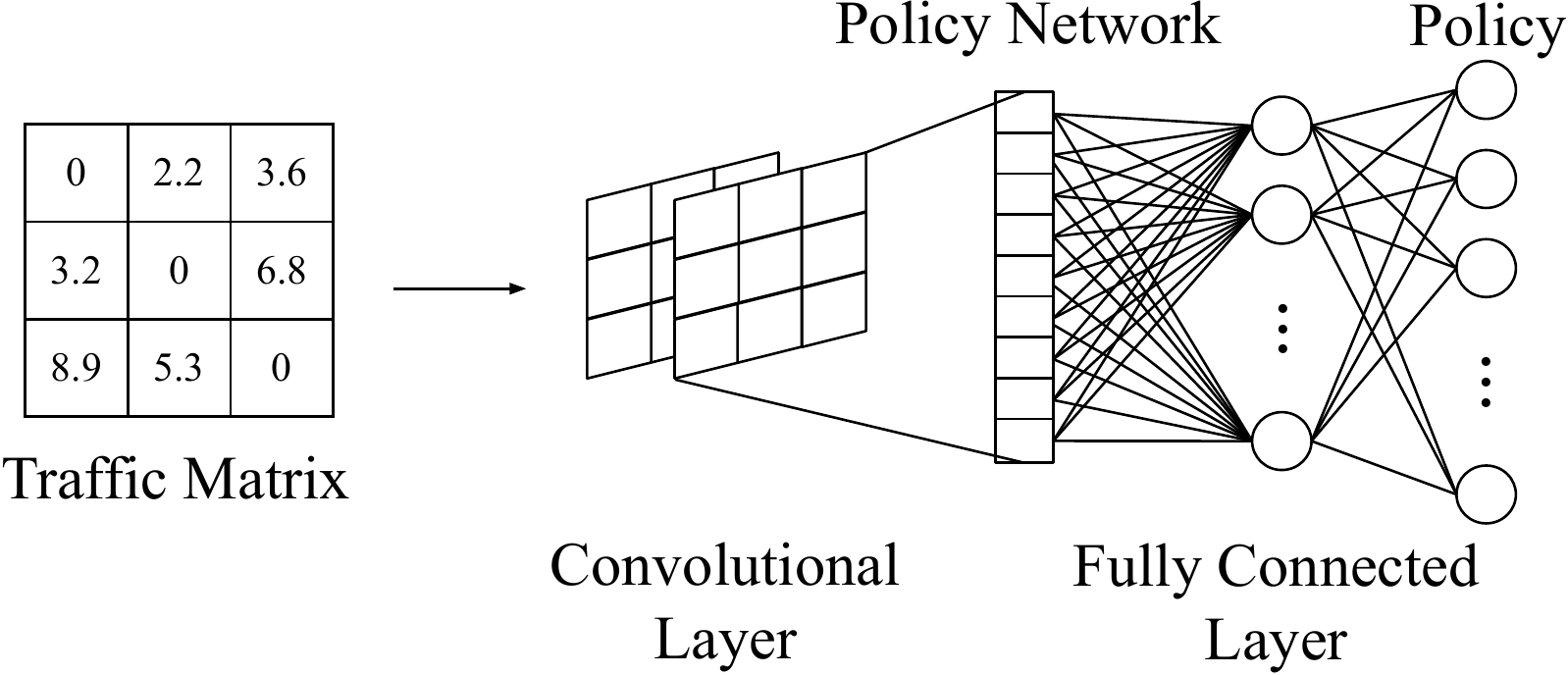}
\caption{Policy network architecture.}
\label{fig:NN}
\end{figure}

The critical flow selection policy is represented by a neural network. This policy network takes a state $s_t =$ $TM_t$ as an input as described above and outputs a probability distribution $\pi(a_t|s_t)$ over all available actions. Figure \ref{fig:NN} shows the architecture of the policy network (details in Section \ref{sec:Implementation}). Since $K$ different actions are sampled for each state $s_t$ and their order does not matter, we define a solution $a_{t_K} = (a_t^1, a_t^2, ..., a_t^K)$ as a combination of $K$ sampled actions. For selecting a solution $a_{t_K}$ with a given state $s_t$, a stochastic policy $\pi(a_{t_K}|s_t)$ parameterized by $\theta$ can be approximated as follows\footnote{To select $K$ distinct actions, we do the action sampling without replacement. The right side of Eq. (\ref{eq:production}) is the solution probability when sampling with replacement, we use Eq. (\ref{eq:production}) to approximate the probability of the solution $a_{t_K}$ given a state $s_t$ for simplicity.}:
\begin{equation}
\pi(a_{t_K}|s_t; \theta) \approx \prod_{i=1}^K\pi(a^i_t|s_t; \theta). \label{eq:production}
\end{equation}
The goal of training is to maximize the network performance over various traffic matrices, i.e., maximize the expected reward $E[r_t]$. Thus, we optimize $E[r_t]$ by gradient ascend, using REINFORCE algorithm with a baseline $b(s_t)$. The policy parameter $\theta$ is updated according to the following equation:
\begin{equation}
\theta \gets \theta + \alpha\sum_t\nabla_{\theta}log\pi(a_{t_K}|s_t;\theta)(r_t - b(s_t)), \label{eq:gradient}
\end{equation}
where $\alpha$ is the learning rate for the policy network. A good baseline $b(s_t)$ reduces gradient variance and thus increases speed of learning. In this paper, we use an average reward for each state $s_t$ as the baseline. ($r_t - b(s_t)$) indicates how much better a specific solution is compared to the "average solution" for a given state $s_t$ according to the policy. Intuitively, Eq.\eqref{eq:gradient} can be explained as follows. If $(r_t - b(s_t))$ is positive, $\pi({a_{t_K}}|s_t;\theta)$ (i.e., the probability of the solution $a_{t_K}$) is increased by updating the policy parameters $\theta$ in the direction $\nabla_{\theta}log\pi({a_{t_K}}|s_t;\theta)$ with a step size of $\alpha(r_t - b(s_t))$. Otherwise, the solution probability is decreased. The net effect of Eq. (\ref{eq:gradient}) is to reinforce actions that empirically lead to better rewards.

To ensure that the RL agent explores the action space adequately during training to discover good policies, the entropy of the policy $\pi$ is added to Eq. (\ref{eq:gradient}). This technique improves the exploration by discouraging premature convergence to suboptimal deterministic policies \cite{A3C}. Then, Eq(\ref{eq:gradient}) is modified to the following equation:
\begin{equation}
\begin{split}
\theta \gets \theta + \alpha\sum_t(\nabla_{\theta}log\pi(a_{t_K}|s_t;\theta)(r_t - b(s_t)) \\
+ \beta\nabla_{\theta}H(\pi(\cdot|s_t;\theta))), \label{eq:gradient+entropy}
\end{split}
\end{equation}
where $H$ is the entropy of the policy (the probability distribution over actions). The hyperparameter $\beta$ controls the strength of the entropy regularization term. Algorithm \ref{alg:training_algorithm} shows the pseudo-code for the training algorithm.

\begin{algorithm}[t]
\caption{Training Algorithm}
\begin{algorithmic} \label{alg:training_algorithm}
\STATE Initialize $\theta$, $v = \{\}$ (keep track the sum of rewards for each state), $n = \{\}$ (keep track the visited count of each state)
\FOR{each iteration}
\STATE $\Delta\theta \gets 0$
\STATE $\{s_t\} \gets$ Sample a batch of states with size $B$
\FOR{$t = 1, ..., B$}
\STATE Sample a solution $a_{t_K}$ according to policy $\pi(a_{t_K}|s_t)$
\STATE Receive reward $r_t$
\IF{$s_t \in v$ and $s_t \in n$}
\STATE $b(s_t) = \frac{v[s_t]}{n[s_t]}$ (average reward for state $s_t$)
\ELSE 
\STATE $b(s_t) = 0$, $v[s_t] = 0$, $n[s_t] = 0$
\ENDIF
\ENDFOR
\FOR{$t = 1, ..., B$}
\STATE $\Delta\theta \gets \Delta\theta + \alpha(\nabla_{\theta}log\pi(a_{t_K}|s_t; \theta)(r_t - b(s_t)) + \textrm{\qquad \qquad} \beta\nabla_{\theta}H(\pi(\cdot|s_t; \theta)))$
\STATE $v[s_t] = v[s_t] + r_t$
\STATE $n[s_t] = n[s_t] + 1$ 
\ENDFOR
\STATE $\theta \gets \theta + \Delta\theta$
\ENDFOR
\end{algorithmic}
\end{algorithm}

\section{Rerouting Critical Flows}  \label{sec:rerouting}

In this section, we describe how to reroute the selected critical flows to  balance link  utilization  of  the  network.

\subsection{Notations}

\vspace{1mm}
\noindent\begin{tabular}{lp{.78\linewidth}}
$G(V,E)$ & network with nodes $V$ and directed edges $E$ ($|V|=N,|E|=M$). \\
$c_{i,j}$ & the capacity of link $\langle i,j \rangle$ ($\langle i,j \rangle \in E$). \\
$l_{i,j}$ & the traffic load on link $\langle i,j \rangle$ ($\langle i,j \rangle \in E$). \\
$D^{s,d}$ & the traffic demand from source $s$ to destination $d$ ($s,d\in V$, $s \neq d$). \\
$\sigma^{s,d}_{i,j}$ & the percentage of traffic demand from source $s$ to destination $d$ routed on link $\langle i,j \rangle$ ($s,d\in V, s \ne d, \langle i,j \rangle \in E$, $\langle s,d \rangle \in f_K$).\\
\end{tabular}
\vspace{1mm}

\subsection{Explicit Routing For Critical Flows}

By default, traffic is distributed according to ECMP routing. We reroute the small set of critical flows (i.e., $f_K$) by conducting explicit routing optimization for these critical flows $\langle s,d \rangle \in f_K$. 

The critical flow rerouting problem can be described as the following. Given a network $G(V,E)$ with the set of traffic demands $D^{s,d}$ for the selected critical flows ($\forall \langle s,d \rangle \in f_K$) and the background link load $\{\bar{l}_{i,j}\}$ contributed by the remaining flows using the default ECMP routing, our objective is to obtain the optimal explicit routing ratios $\{\sigma^{s,d}_{i,j}\}$ for each critical flow, so that the maximum link utilization $U$ is minimized.

To search all possible under-utilized paths for the selected critical flows, we formulate the rerouting problem as an optimization as follows.

\vspace{0.1cm}
\hrule
\vspace{0.1cm}
\begin{subequations}
\begin{equation}
minimize\quad U +  \epsilon \cdot \sum\limits_{\langle i,j \rangle \in E}\sum\limits_{\langle s,d \rangle \in f_K}\sigma^{s,d}_{i,j}                \label{eq:objective}
\end{equation}
subject to
\begin{equation}
l_{i,j} = \sum\limits_{\langle s,d \rangle \in f_K} \sigma^{s,d}_{i,j} \cdot D^{s,d} + \bar{l}_{i,j} \qquad i,j : \langle i,j \rangle \in E	\label{eq:link_load_hybrid}
\end{equation}
\begin{equation}
l_{i,j} \le c_{i,j} \cdot U \qquad i,j : \langle i,j \rangle \in E          \label{eq:capacity_constraint}
\end{equation}
\begin{equation}
\begin{aligned}
\sum\limits_{k: \langle k,i \rangle \in E} \sigma^{s,d}_{k,i} - \sum\limits_{k: \langle i,k \rangle \in E} \sigma^{s,d}_{i,k} = \left\{
\begin{aligned}
-1 \quad \textrm{if $i=s$} \\
1 \quad \textrm{if $i=d$} \\
0 \quad \textrm{otherwise} \\
\end{aligned}
\right.	\\
\qquad i \in V, \, s, d : \langle s,d \rangle \in f_K  \end{aligned}\label{eq:explicit_routing_flow_constraint}
\end{equation}
\begin{equation}
0 \le \sigma^{s,d}_{i,j} \le 1 \qquad s, d : \langle s,d \rangle \in f_K, \, i,j : \langle i,j \rangle \in E \label{eq:link_ratio}
\end{equation}
\end{subequations}
\hrule
\vspace{0.1cm}

$\epsilon \cdot \sum\limits_{\langle i,j \rangle \in E}\sum\limits_{\langle s,d \rangle \in f_K}\sigma^{s,d}_{i,j}$ in (\ref{eq:objective}) is needed because otherwise the optimal solution may include unnecessarily long paths as long as they avoid the most congested link, where $\epsilon$ ($\epsilon>0$) is a sufficiently small constant to ensure that the minimization of $U$ takes higher priority \cite{916782}. (\ref{eq:link_load_hybrid}) indicates the traffic load on link $\langle i,j \rangle$ contributed by the traffic demands routed by the explicit routing and the traffic demands routed by the default ECMP routing. (\ref{eq:capacity_constraint}) is the link capacity utilization constraint. (\ref{eq:explicit_routing_flow_constraint}) is the flow conservation constraint for the selected critical flows.

By solving the above LP problem using LP solvers (such as Gurobi \cite{gurobi}), we can obtain the optimal explicit routing solution for selected critical flows $\{\sigma^{s,d}_{i,j}\}$ $(\forall \langle s,d \rangle \in f_K)$. Then, the SDN controller installs and updates flow entries at the switches accordingly. 

\section{Evaluation}
In this section, a series of simulation experiments are conducted using real-world network topologies to evaluate the performance of CFR-RL and show its effectiveness by comparing it with other rule-based heuristic schemes.

\subsection{Evaluation Setup}

\subsubsection{Implementation} \label{sec:Implementation}
The policy neural network consists of three layers. The first layer is a convolutional layer with 128 filters. The corresponding kernel size is $3 \times 3$ and the stride is set to 1. The second layer is a fully connected layer with 128 neurons. The activation function used for the first two layers is Leaky ReLU \cite{leaky_relu}. The final layer is a fully connected linear layer (without activation function) with $N*(N-1)$ neurons corresponding to all possible critical flows. The softmax function is applied upon the output of final layer to generate the probabilities for all available actions. The learning rate $\alpha$ is initially configured to 0.001 and decays every 500 iterations with a base of 0.96 until it reaches the minimum value 0.0001. Additionally, the entropy factor $\beta$ is configured to be 0.1. We found that the set of above hyperparameters is a good trade-off between performance and computational complexity of the model (details in Section \ref{sec: hyperparameters}). Thus, we fixed them throughout our experiments. The results in the following experiments show CFR-RL works well on different network topologies with a single set of fixed hyperparameters. This architecture is implemented using TensorFlow \cite{tensorflow}.

\begin{table}[htb]
    \centering
    \caption{ISP networks used in evaluation}
    \label{tbl:Statistics}
    \begin{tabular}{|c|c|c|c|}
        \hline
        Topology & Nodes & Directed Links & Pairs\\
        \hline
        Abilene & 12 & 30 & 132\\
        \hline
        EBONE (Europe) & 23 & 74 & 506\\
        \hline
        Sprintlink (US) & 44 & 166 & 1892\\
        \hline
        Tiscali (Europe) & 49 & 172 & 2352\\
        \hline
    \end{tabular}
\end{table}

\subsubsection{Dataset}
In our evaluation, we use four real-world network topologies including Abilene network and 3 ISP networks collected by ROCKETFUEL \cite{spring2002measuring}. 
The number of nodes and directed links of the networks are listed in Table \ref{tbl:Statistics}. For the Abilene network, the measured traffic matrices and network topology information (such as link connectivity, weights, and capacities) are available in \cite{abilene_tm}. 
Since Abilene traffic matrices are measured every 5 minutes, there are a total of 288 traffic matrices each day. To evaluate the performance of CFR-RL, we choose a total 2016 traffic matrices in the first week (starting from Mar. 1st 2004) as our dataset. For ROCKETFUEL topologies, the link costs are given while the link capacities are not provided. Therefore, we infer the link capacities as the inverse of link costs, which is based on the default link cost setting in Cisco routers. In other words, the link costs are inversely proportional to the link capacities.
This approach is commonly adopted in literature \cite{zhang2014dynamic, zhang2014hybrid_routing, kodialam2008oblivious}. 
Besides, since traffic matrices are also unavailable for the ISP networks from ROCKETFUEL, we use a traffic matrix generation tool \cite{TMgen} to generate 50 synthetic exponential traffic matrices and 50 synthetic uniform traffic matrices for each network. Unless otherwise noted, we use a random sample of 70\% of our dataset as a training set for CFR-RL, and use the remaining 30\% as a test set for testing all schemes.
\subsubsection{Parallel Training} 
\label{sec:Parallel_training}
To speed up training, we spawn multiple actor agents in parallel, as suggested by \cite{A3C}. CFR-RL uses 20 actor agents by default. Each actor agent is configured to experience a different subset of the training set. Then, these agents continually forward their (state, action, advantage (i.e, $r_t - b(s_t)$)) tuples to a central learner agent, which aggregates them to train the policy neural network. The central learner agent performs a gradient update using Eq(\ref{eq:gradient+entropy}) according to the received tuples, then sends back the updated parameters of the policy network to the actor agents. The whole process can happen asynchronously among all agents. We use 21 CPU cores to train CFR-RL (i.e., one core (2.6GHz) for each agent).
\subsubsection{Metrics} \label{sec: metrics}

\noindent {(1) Load Balancing Performance Ratio:}
To demonstrate the load balancing performance of the proposed CFR-RL scheme, a load balancing performance ratio is applied and defined as follows:
\begin{equation}
PR_U = \frac{U^{\textrm{optimal}}}{U^{\textrm{CFR-RL}}}, \label{eq:load_balancing_performance_ratio}
\end{equation}
where ${U^{\textrm{optimal}}}$ is the maximum link utilization achieved by an optimal explicit routing for all flows\footnote{The corresponding LP formulation is similar to \eqref{eq:objective}, except that the objective becomes obtaining the optimal explicit ratios $\{\sigma^{s,d}_{i,j}\}$ for all flows. Note that the background link load $\{\bar{l}_{i,j}\}$ would be 0 for this problem.}. $PR_U = 1$ means that the proposed CFR-RL achieves load balancing as good as the optimal routing. A lower ratio indicates that the load balancing performance of CFR-RL is farther away from that of the optimal routing.

\noindent {(2) End-to-end Delay Performance Ratio:}
To model and measure end-to-end delay in the network, we define the overall end-to-end delay in the network as $\Omega = \sum\limits_{\langle i,j \rangle \in E}(\frac{l_{i.j}}{c_{i,j}-l_{i,j}})$ as described in \cite{zhang2012optimizing}.
Then, an end-to-end delay performance ratio is defined as follows:
\begin{equation}
PR_{\Omega} = \frac{\Omega^{\textrm{optimal}}}{\Omega^{\textrm{CFR-RL}}}, \label{eq:delay_performance_ratio}
\end{equation}
where ${\Omega^{\textrm{optimal}}}$ is the minimum end-to-end delay achieved by an optimal explicit routing for all flows with an objective\footnote{The objective of this LP problem is to obtain the optimal explicit routing ratios $\{\sigma^{s,d}_{i,j}\}$ for all flows, such that $\Omega$ is minimized.} to minimize the end-to-end delay $\Omega$. Note that the rerouting solution for selected critical flows is still obtained by solving \eqref{eq:objective}. The higher $PR_{\Omega}$, the better end-to-end delay performance achieved by CFR-RL. $PR_{\Omega} = 1$ means that the proposed CFR-RL achieves the minimum end-to-end delay as the optimal routing.

\noindent {(3) Rerouting Disturbance:}
To measure the disturbance caused by rerouting, we define rerouting disturbance as the percentage of total rerouted traffic\footnote{Although partial of traffic flows might still be routed along the original ECMP paths, updating routing at the switches might cause packets drop or out-of-order. Thus, we still consider this amount of traffic as rerouting traffic.} for a given traffic matrix, i.e., 
\begin{equation}
RD = \frac{\sum\limits_{\langle s,d \rangle \in f_K} D^{s,d}}{\sum\limits_{s,d\in V,s \ne d} D^{s,d}}, \label{eq:rerouting_disturbance}
\end{equation}
where $\sum\limits_{\langle s,d \rangle \in f_K} D^{s,d}$ is the total traffic of selected critical flows that need to be rerouted and $\sum\limits_{s,d\in V,s \ne d} D^{s,d}$ is the total traffic of all flows. The smaller $RD$, the less disturbance caused by rerouting.

\subsubsection{Rule-based Heuristics}

For comparison, we also evaluate two rule-based heuristics as the following:
\begin{enumerate}
\item{\textbf{Top-K}:} selects the $K$ largest flows from a given traffic matrix in terms of demand volume. This approach is based on the assumption that flows with larger traffic volumes would have a dominant impact to network performance.
\item{\textbf{Top-K Critical}:} similar to Top-$K$ approach, but selects the $K$ largest flows from the most congested links. This approach is based on the assumption that flows traversing the most congested links would have a dominant impact to network performance.
\end{enumerate}

\subsection{Evaluation} \label{sec:evaluation}

\subsubsection{Critical Flows Number} \label{sec:critical_flows_num}

\begin{figure}[htb]
    \centering
        \includegraphics[width=0.9\linewidth]{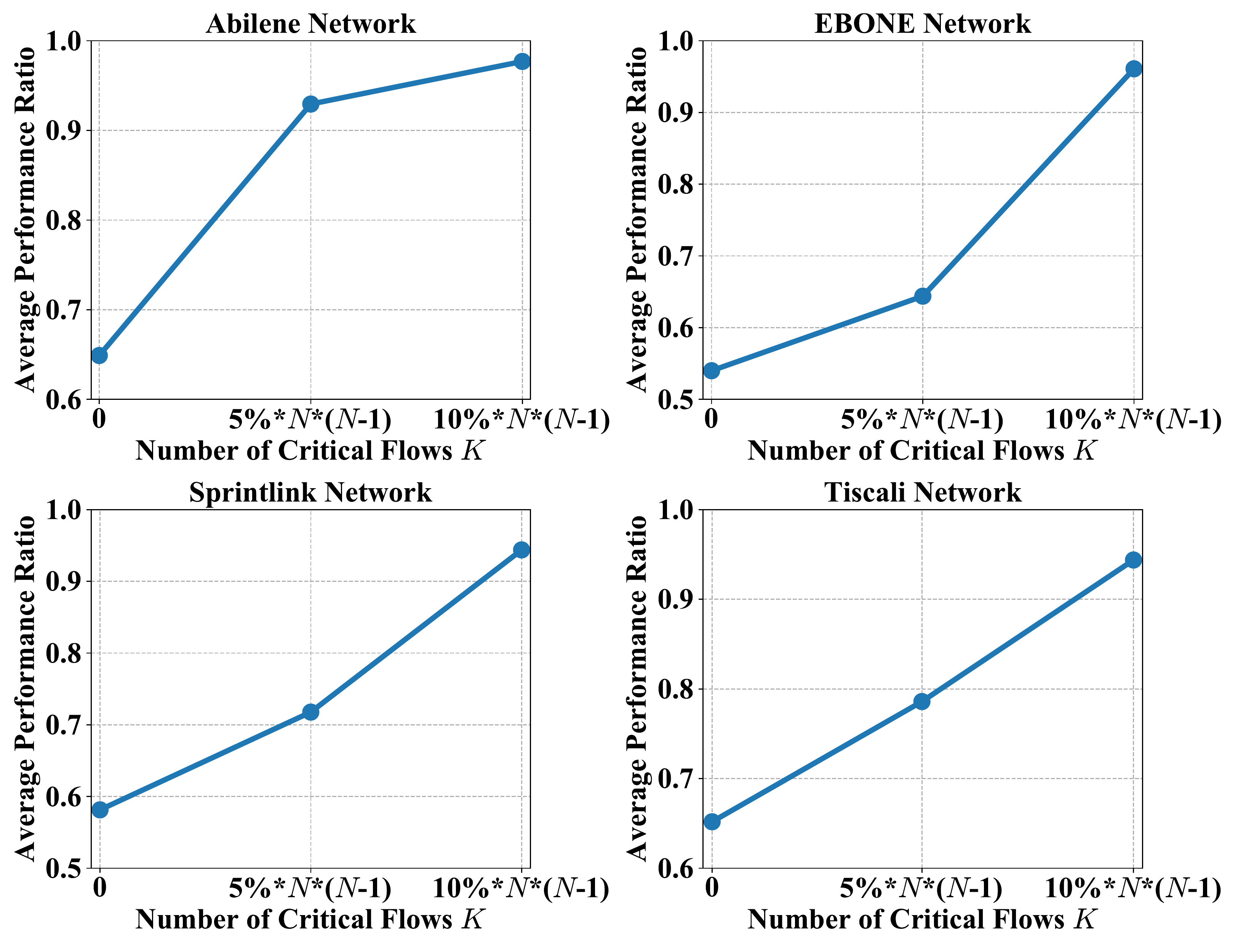}
        \caption{Average load balancing performance ratio of CFR-RL with increasing number of critical flows $K$ on the four networks.}
        \label{exp:parameter}
\end{figure}

We conduct a series of experiments with different number of critical flows selected, and fix other parameters throughout the experiments.

Figure \ref{exp:parameter} shows the average load balancing performance ratio achieved by CFR-RL with increasing number of critical flows $K$. The initial value with $K$ = 0 represents the default ECMP routing. The results indicate that there is a considerable room for further improvement when flows are routed by ECMP. The sharp increases in the average load balancing performance ratio for all four networks shown in Fig. \ref{exp:parameter} indicates that CFR-RL is able to achieve near-optimal load balancing performance by rerouting only 10\% flows. As a result, network disturbance would be much reduced compared to rerouting all flows as traditional TE. For the subsequent experiments, we set $K = 10\%*N*(N-1)$ for each network.

\subsubsection{Performance Comparison} \label{sec:performance_comparison}
\begin{figure}[tb]
    \centering
        \includegraphics[width=1\linewidth]{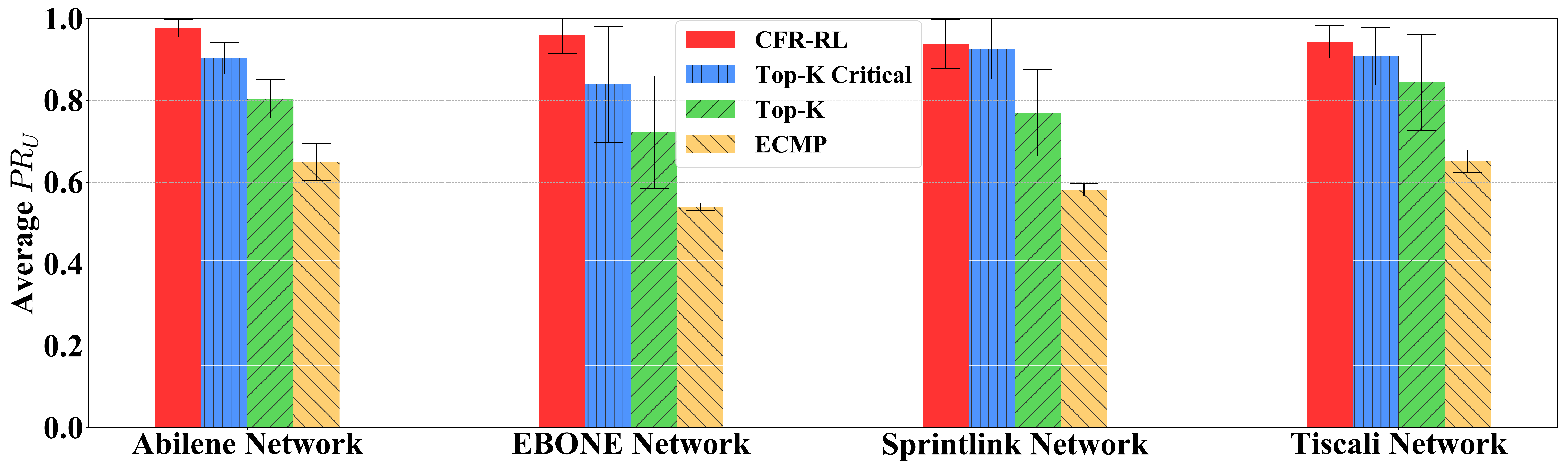}
        \caption{Comparison of average load balancing performance ratio where error bars span $\pm$ one standard deviation from the average on the entire test set of the four networks.}
        \label{exp:bar}
\end{figure}

\begin{figure}[htb]
    \centering
        \includegraphics[width=1\linewidth]{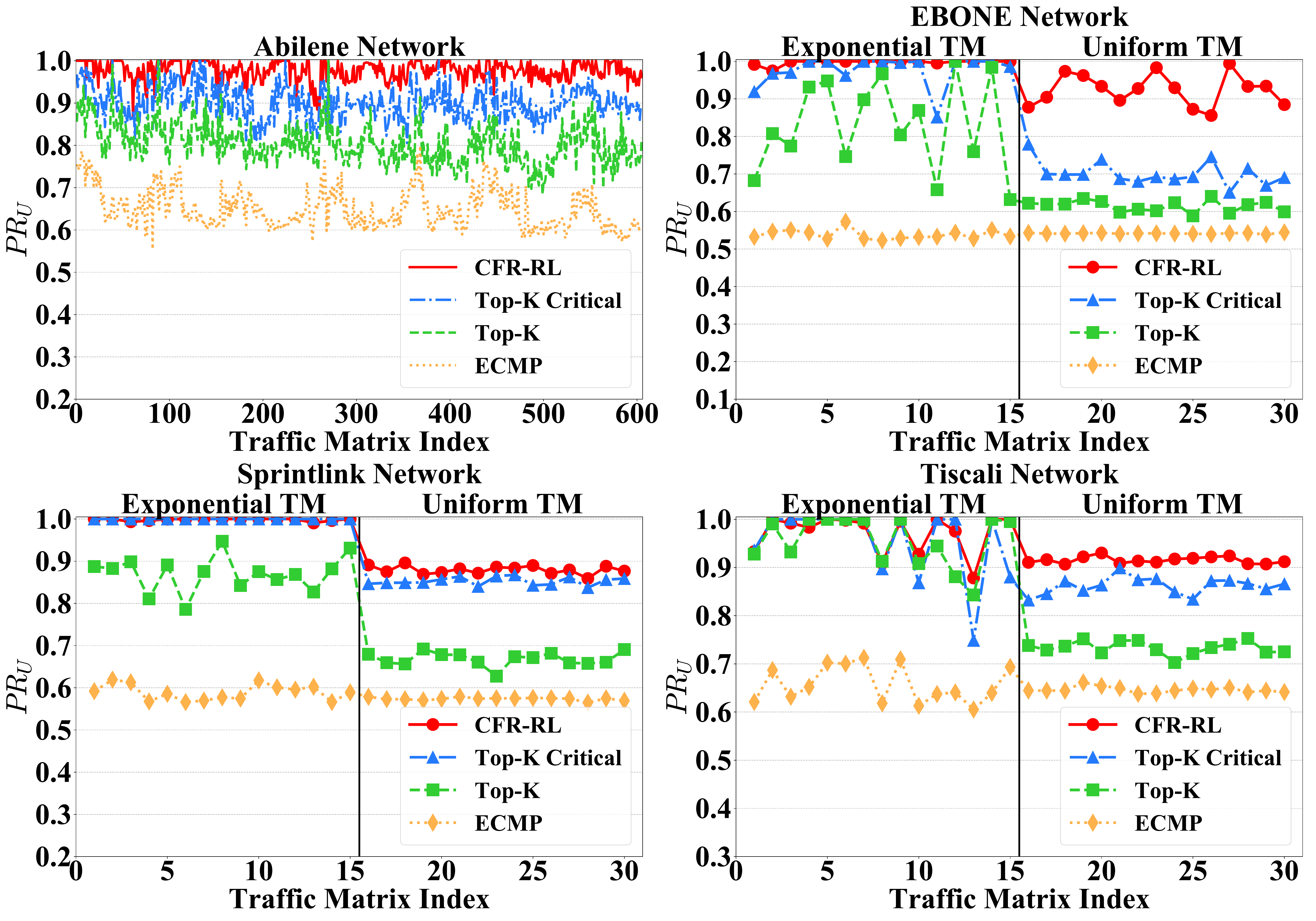}
        \caption{Comparison of load balancing performance in the four networks on each test traffic matrix.}
        \label{exp:tm}
\end{figure}

\begin{figure}[htb]
    \centering
        \includegraphics[width=1\linewidth]{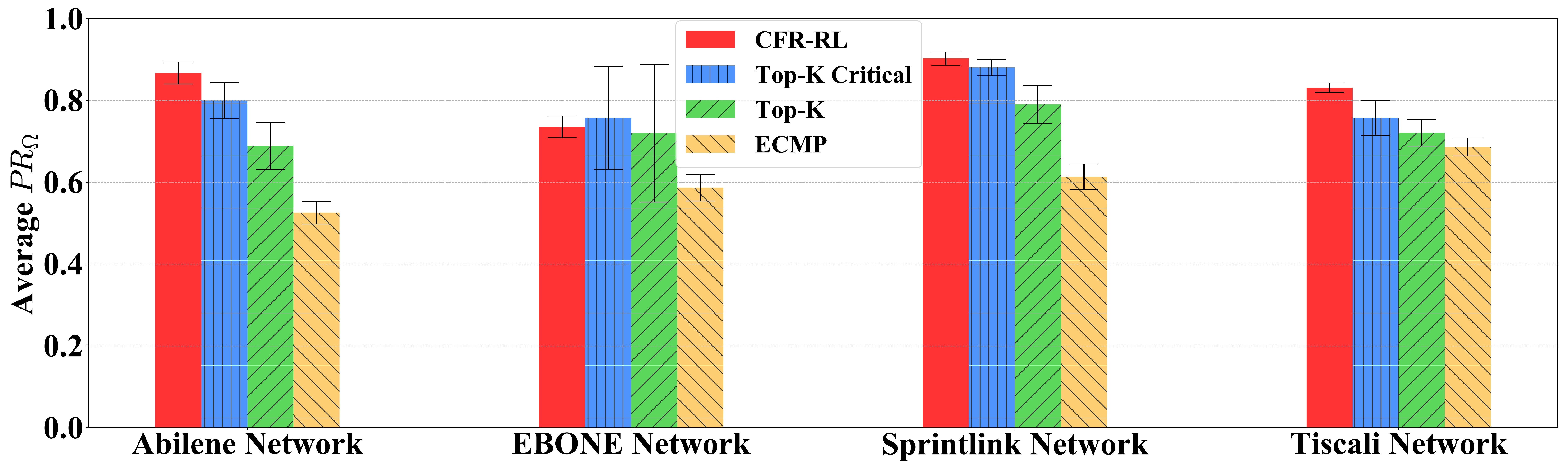}
        \caption{Comparison of average end-to-end delay performance ratio where error bars span $\pm$ one standard deviation from the average on the entire test set of the four networks.}
        \label{exp:bar_d}
\end{figure}

\begin{figure}[htb]
    \centering
        \includegraphics[width=1\linewidth]{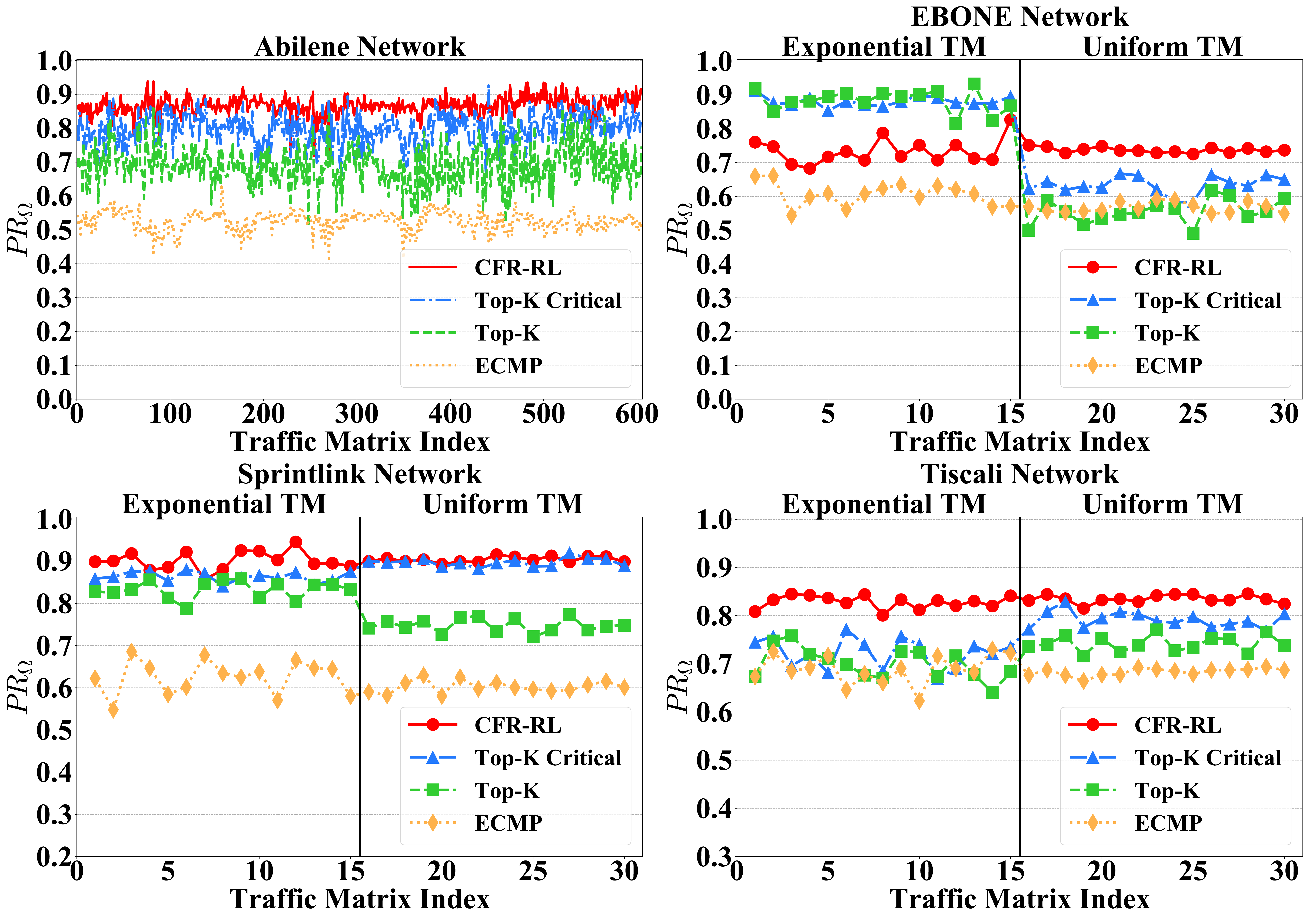}
        \caption{Comparison of end-to-end delay performance in the four networks on each test traffic matrix.}
        \label{exp:tmd}
\end{figure}

\begin{table*}[ht]
    \centering
    \caption{Comparison of average rerouting disturbance}
    \label{tbl:Traffic}
    \begin{tabular}{|c|c|c|c|}
        \hline
        \textbf{Topology} & \ \textbf{CFR-RL} & \textbf{Top-K Critical} & \textbf{Top-K}\\
        \hline
        Abilene & 21.3\% & 32.7\% & 42.9\% \\
        \hline
        EBONE (Exponential / Uniform) & 11.2\% / 10.0\% & 25.9\% / 11.5\% & 32.9\% / 11.7\%\\
        \hline
        Sprintlink (Exponential / Uniform) & 11.3\% / 10.1\% & 23.6\% / 13.8\% & 33.2\% / 14.6\%\\
        \hline
        Tiscali (Exponential / Uniform) & 11.2\% / 10.0\% & 24.5\% / 12.0\% & 32.7\% / 12.2\%\\
        \hline
    \end{tabular}
\end{table*}

\begin{figure}[htb]
    \centering
        \includegraphics[width=1\linewidth]{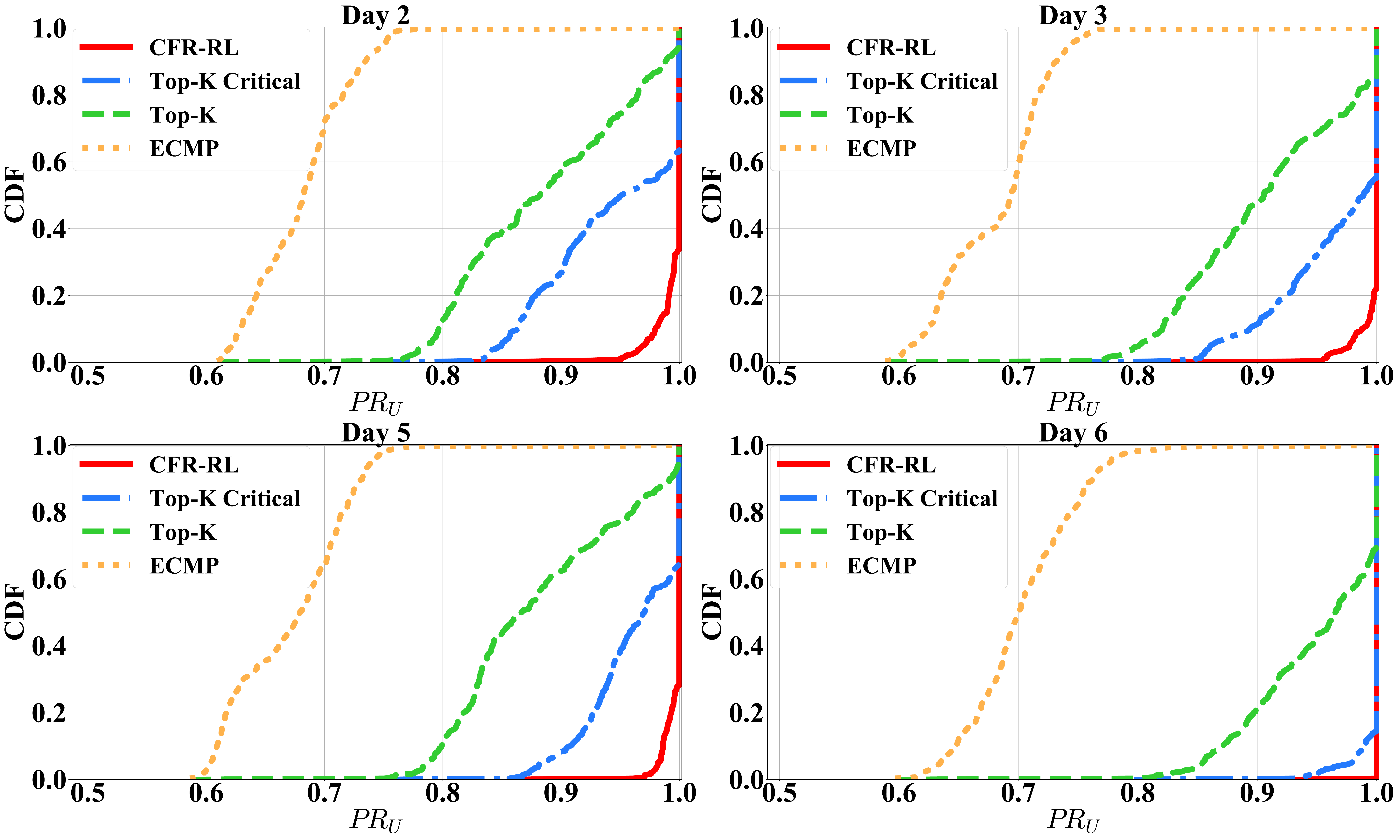}
        \caption{Comparison of load balancing performance ratio in CDF with the traffic matrices from Tuesday, Wednesday, Friday and Saturday in week 2.}
        \label{exp:abilene2}
\end{figure}

\begin{figure}[htb]
    \centering
        \includegraphics[width=1\linewidth]{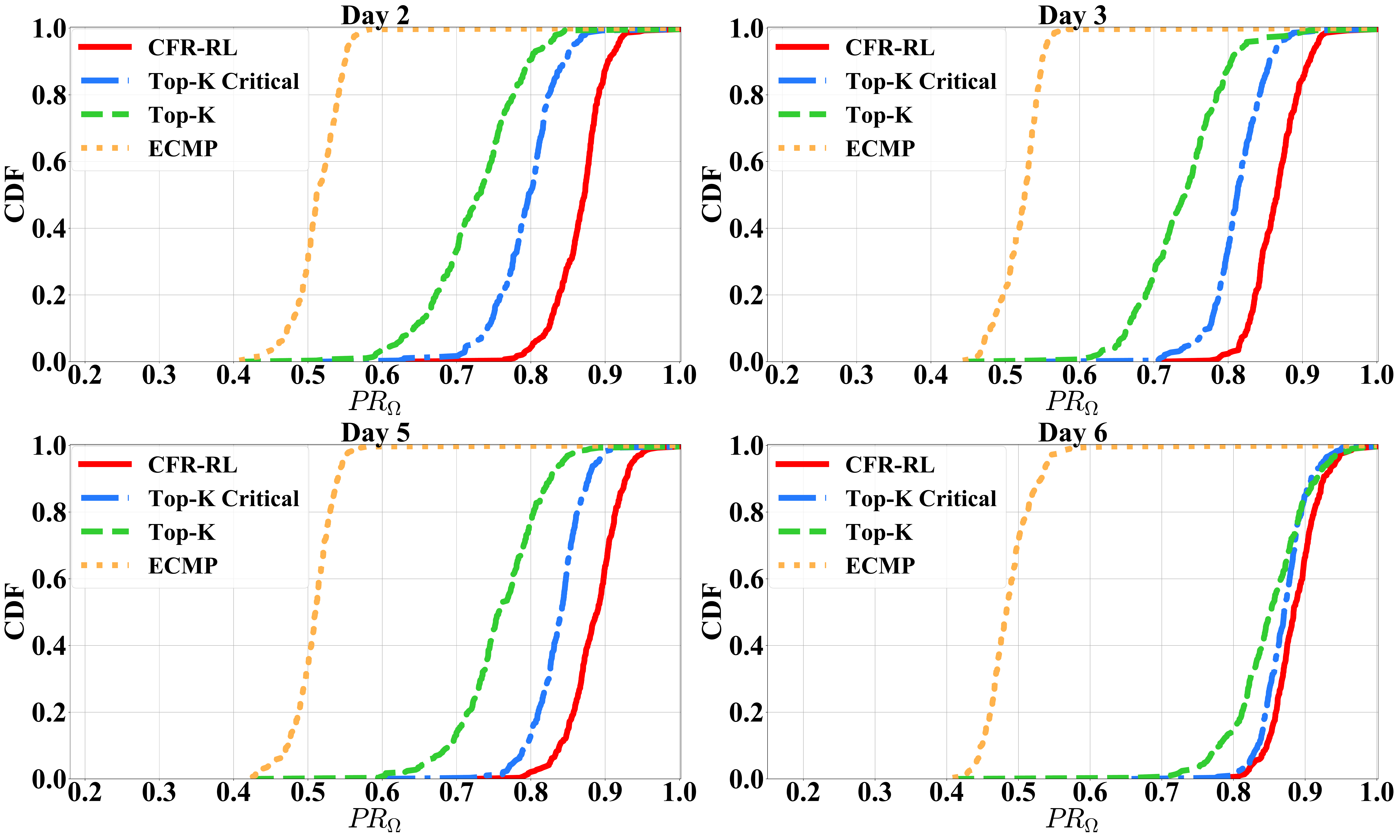}
        \caption{Comparison of end-to-end delay performance ratio in CDF with the traffic matrices from Tuesday, Wednesday, Friday and Saturday in week 2.}
        \label{exp:abilene2d}
\end{figure}

\begin{figure}[htb]
    \centering
        \includegraphics[width=1\linewidth]{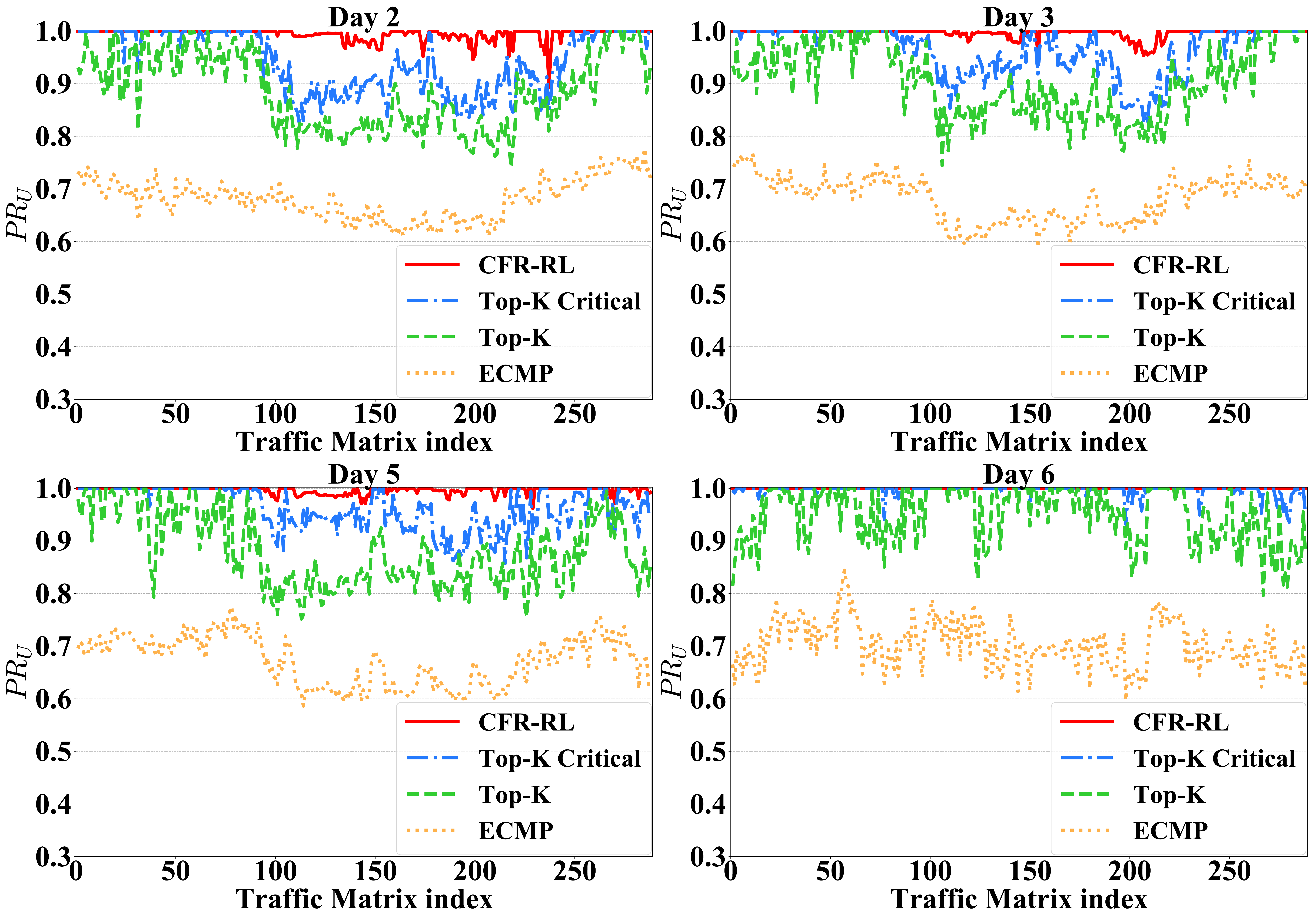}
        \caption{Comparison of load balancing performance ratio with the traffic matrices from Tuesday, Wednesday, Friday and Saturday in week 2.}
        \label{exp:abilene2tm}
\end{figure}

\begin{figure}[htb]
    \centering
        \includegraphics[width=1\linewidth]{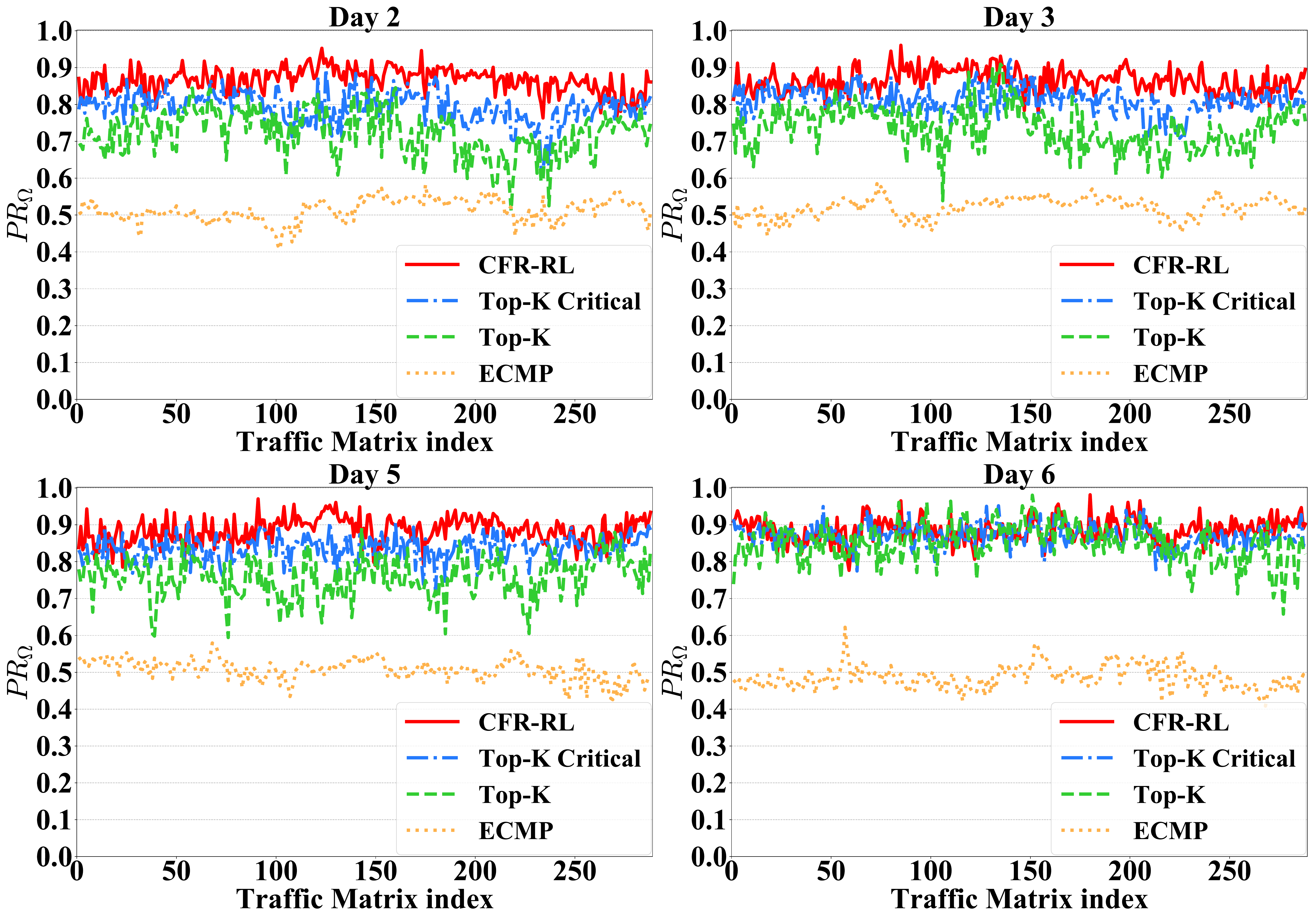}
        \caption{Comparison of end-to-end delay performance ratio with the traffic matrices from Tuesday, Wednesday, Friday and Saturday in week 2.}
        \label{exp:abilene2tmd}
\end{figure}

For comparison, we also calculate the performance ratios and rerouting disturbances for Top-K, Top-K critical, and ECMP according to Eqs. \eqref{eq:load_balancing_performance_ratio}, \eqref{eq:delay_performance_ratio} and \eqref{eq:rerouting_disturbance}. Figure \ref{exp:bar} shows the average load balancing performance ratio that each scheme achieves on the entire test set of the four networks. Figure \ref{exp:tm} shows the load balancing performance ratio on each individual traffic matrix for the four networks. Note that the first 15 traffic matrices in Figs. \ref{exp:tm}(b)-\ref{exp:tm}(d) are generated by an exponential model and the remaining 15 traffic matrices are generated by an uniform model. CFR-RL performs significantly well in all networks. For example, for the Abilene network, CFR-RL improves load balancing performance by about 32.8\% compared to ECMP, and by roughly 7.4\% compared to Top-K critical. For the EBONE network, CFR-RL outperforms Top-K critical with an average 12.2\% load balancing performance improvement. For Sprintlink and Tiscali networks, CFR-RL performs slightly better than Top-K critical by 1.3\% and 3.5\% on average, respectively. 
Moreover, Figure \ref{exp:bar_d} shows the average end-to-end delay performance ratio that each scheme achieves on the entire test set of the four networks. Figure \ref{exp:tmd} shows the end-to-end delay performance ratio on each test traffic matrix for the four networks. It is worth noting that the rerouting solution for selected critical flows is still obtained by solving \eqref{eq:objective} (i.e., minimize maximum link utilization), though the end-to-end delay performance is evaluated\footnote{For the Abilene network, the real traffic demands in the measured traffic matrices collected in \cite{abilene_tm} are relatively small, and thus the corresponding end-to-end delay would be very small. To effectively compare end-to-end delay performance of each scheme, we multiply each demand $D^{s,d}$ in a real traffic matrix $TM_t$ by $\frac{0.9}{U_t^{\textrm{ECMP}}}$, where $U_t^{\textrm{ECMP}}$ is the maximum link utilization achieved by ECMP routing on the traffic matrix $TM_t$.} for each scheme. By effectively selecting and rerouting critical flows to balance link utilization of the network, CFR-RL outperforms heuristic schemes and ECMP in terms of end-to-end delay in all networks except the EBONE network. In the EBONE network, heuristic schemes performs better with the exponential traffic model. It is possible that rerouting the elephant flows selected by heuristic schemes further balances load on non-congested links and results in achieving smaller end-to-end delay.
In addition, Tab. \ref{tbl:Traffic} shows the average rerouting disturbance, i.e., the average percentage of total traffic rerouted by each scheme (except ECMP) for the four networks. CFR-RL greatly reduces network disturbance by rerouting at most 21.3\%, 11.2\%, 11.3\%, and 11.2\% of total traffic on average for the four networks, respectively. In contrast, Top-K critical reroutes 11.4\% more traffic for the Abilene network and 14.7\%, 12.3\%, and 13.3\% more traffic for the EBONE, Sprintlink, and Tiscali networks (for exponential traffic matrices). Top-K performs even worse by rerouting more than 42\% of total traffic on average for the Abilene network and 32\%, 33\%, and 32\% of total traffic on average for the other three networks (for exponential traffic matrices). It is worth noting that there are no elephant flows in uniform traffic matrices shown in Fig. \ref{exp:tm}(b)-\ref{exp:tm}(d). Thus, all three schemes reroute similar amount of traffic for uniform traffic matrices. However, CFR-RL is still able to perform slightly better than the two rule-based heuristics. Overall, the above results indicate that CFR-RL is able to achieve near-optimal load balancing performance and greatly reduce end-to-end delay and network disturbance by smartly selecting a small number of critical flows for each given traffic matrix and effectively rerouting the corresponding small amount of traffic.

As shown in Figs. \ref{exp:tm}(b)-\ref{exp:tm}(d), Top-K critical performs well with the exponential traffic model. However, its performance is degraded with the uniform traffic model. One possible reason for the performance degradation of Top-K critical is that all links in the network are relatively saturated under the uniform traffic model. Alternative underutilized paths are not available for the critical flows selected by Top-K critical. In other words, there is no much room for rerouting performance improvement by only considering the elephant flows traversing the most congested links. Thus, fixed-rule heuristics are unable to guarantee their performance, showing that their design assumptions are invalid. In contrast, CFR-RL performs consistently well under various traffic models. 

\subsubsection{Generalization}      \label{sec:generalization}

In this series of experiments, we trained CFR-RL on the traffic matrices from the first week (starting from Mar. 1st 2004) and evaluate it for each day of the following week (starting from Mar. 8th 2004) for the Abilene network. We only present the results for day 2, day 3, day 5 and day 6, since the results for other days are similar. Figures \ref{exp:abilene2} and \ref{exp:abilene2d} show the full CDFs of two types of performance ratio for these 4 days. Figures \ref{exp:abilene2tm} and \ref{exp:abilene2tmd} show the load balancing and end-to-end delay performance ratios on each traffic matrix of these 4 days, respectively. The results show that CFR-RL still achieves above 95\% optimal load balancing performance and average 88.13\% end-to-end delay performance, and thus outperforms other schemes on almost all traffic matrices. The load balancing performance of CFR-RL degrades on several outlier traffic matrices in day 2. There are two possible reasons for the degradation: (1) The traffic patterns of these traffic matrices are different from what CFR-RL learned from the previous week. (2) Selecting $K = 10\%*N*(N-1)$ is not enough for CFR-RL to achieve near-optimal performance on these outlier traffic matrices. However, CFR-RL still performs better than other schemes. Overall, the results indicate that real traffic patterns are relatively stable and CFR-RL generalizes well to unseen traffic matrices for which it was not explicitly trained.

\subsubsection{Training and Inference Time} \label{sec:training_inference_time}

Training a policy for the Abilene network took approximately 10,000 iterations, and the time consumed for each iteration is approximately 1 second. As a result, the total training time for Abilene network is approximately 3 hours. Since the EBONE network is relatively larger, it took approximately 60,000 iterations to train a policy. Then, the total training time for EBONE network is approximately 16 hours. For larger networks like Sprintlink and Tiscali, the solution space is even larger. Thus, more iterations (e.g., approximately 90,000 and 100,000 iterations) should be taken to train a good policy, and each iteration takes approximately 2 seconds. Note that this cost is incurred offline and can be performed infrequently depending on environment stability. The policy neural network as described in Section \ref{sec:Implementation} is relatively small. Thus, the inference time for the Abilene and EBONE networks are less than 1 second, and they are less than 2 seconds for the Sprintlink and Tiscali networks.

\subsubsection{Hyperparameters} \label{sec: hyperparameters}

\begin{table}[tb]
    \centering 
    \caption{Comparison of average load balancing performance ratio with different sets of hyperparameters} \label{tbl: hyperparameters}
    \begin{threeparttable}
    \begin{tabular}{|c|c|}
        \hline
         & filters / neurons = 128, $\beta = 0.1$ \\
        \hline
        $\alpha = 0.01$ (with decay) & 0.761  \\
        $\mathbf{\alpha = 0.001}$ (with decay) & \textbf{0.970}  \\
        $\alpha = 0.0001$\tnote{*} & 0.963 \\
        \hline
    \end{tabular}
    \begin{tablenotes}
    \item[*] without decay, since the initial learning rate is equal to the minimum learning rate.
    \end{tablenotes}
    \end{threeparttable}
    \vspace{1mm}
    \\(a)
    \bigskip
    
    \begin{tabular}{|c|c|}
        \hline
         & $\alpha = 0.001$ (with decay), $\beta = 0.1$ \\
        \hline
        filters / neurons = 64 & 0.928  \\
        filters / neurons = \textbf{128} & \textbf{0.970}  \\
        filters / neurons = 256 & 0.837  \\
        \hline
    \end{tabular}
    \vspace{1mm}
    \\(b)
    \bigskip
    
    \begin{tabular}{|c|c|}
        \hline
         & filters / neurons = 128, $\alpha = 0.001$ (with decay) \\
        \hline
        $\mathbf{\beta = 0.1}$ & \textbf{0.970}  \\
        $\beta = 0.01$ & 0.958  \\
        \hline
    \end{tabular}
    \vspace{1mm}
    \\(c)
\end{table}

Table \ref{tbl: hyperparameters} shows that how hyperparameters affect the load balancing performance of CFR-RL in the Abilene network. For each set of hyperparameters, we trained a policy for the Abilene network by 10,000 iterations, and then evaluated the average load balancing performance ratio over the whole test set. We only presented the results for the Abilene network, since the results for other network topologies are similar. In Tab. \ref{tbl: hyperparameters}(a), the number of filters in the convolutional layer and neurons in the fully connected layer is fixed to 128 and entropy factor $\beta$ is fixed to 0.1. We compare the performance with different learning rate $\alpha$. The results show that training might become unstable if the initial learning rate is too large (e.g., 0.01), and thus it cannot converge to a good policy. In contrast, training with a smaller learning rate is more stable but might require longer training time to further improve the performance. As a result, we chose $\alpha = 0.001$ to encourage exploration in the early stage of training. We compared the performance with different sizes of filters and neurons in Tab. \ref{tbl: hyperparameters}(b). The results show that too few filters/neurons might restrict the representation that the neural network can learn and thus causes under-fitting. Meanwhile, too many neurons might cause over-fitting, and thus the corresponding policy cannot generalize well to the test set. In addition, more training time is required for a larger neural network. In Tab. \ref{tbl: hyperparameters}(c), the results show that a larger entropy factor encourages exploration and leads to a better performance. Overall, the set of hyperparameters we have chosen is a good trade-off between performance and computational complexity of the model.

\section{Conclusion and Future Work}
With an objective of minimizing the maximum link utilization in a network and reducing disturbance to the network causing service disruption, we proposed CFR-RL, a scheme that learns a critical flow selection policy automatically using reinforcement learning, without any domain-specific rule-based heuristic. CFR-RL selects critical flows for each given traffic matrix and reroutes them to balance link utilization of the network by solving a simple rerouting optimization problem. Extensive evaluations show that CFR-RL achieves near-optimal performance by rerouting only a limited portion of total traffic. In addition, CFR-RL generalizes well to traffic matrices for which it was not explicitly trained.

Yet, there are several aspects that may help improving the solution that we proposed in this contribution. Among them, we are determining how CFR-RL can be updated and improved.  

\medbreak
\noindent {\bf Objectives:}
CFR-RL could be formulated to achieve other objectives. For example, to minimize overall end-to-end delay in the network (i.e., $\Omega = \sum\limits_{\langle i,j \rangle \in E}(\frac{l_{i.j}}{c_{i,j}-l_{i,j}})$) described in Section \ref{sec: metrics}(2), we can define reward $r$ as $1/\Omega$ and reformulate the rerouting optimization problem (\ref{eq:objective}) to minimize $\Omega$. 

Table \ref{tbl:Traffic} shows an interesting finding. Although CFR-RL does not explicitly minimize rerouting traffic, it ends up rerouting much less traffic (i.e., 10.0\%-21.3\%) and performs better than rule-based heuristic schemes by 1.3\%-12.2\%. This reveals that CFR-RL is effectively searching the whole set of candidate flows to find the best critical flows for various traffic matrices, rather than simply considering the elephant flows on the most congested links or in the whole network as rule-based heuristic schemes do. We will consider minimizing rerouting traffic as one of our objectives and investigate the trade-off between maximizing performance and minimizing rerouting traffic. 

\medbreak
\noindent {\bf Scalability:}
Scaling CFR-RL to larger networks is an important direction of our future work. CFR-RL relies on LP to produce reward signals $r$. The LP problem would become complex as the number of critical flows $K$ and the size of a network increase. This would slow down the policy training for larger networks (e.g., the Tiscali network in Section \ref{sec:training_inference_time}), since the time consumed for each iteration would increase. Moreover, the solution space would become enormous for larger networks, and RL has to take more iterations to converge to a good policy. To further speed up training, we can either spawn even more actor agents (e.g., 30) in parallel to allow the system to consume more data at each time step and thus improve exploration \cite{A3C}, or apply GA3C \cite{GA3C} to offload the training to a GPU, which is an alternative architecture of A3C and emphasizes on an efficient GPU utilization to increase the number of training data generated and processed per second. Another possible design to mitigate the scalability issue is adopting SDN multi-controller architectures. Each controller takes care of a subset of routers in a large network, and one CFR-RL agent is running on each SDN controller. The corresponding problem naturally falls into the realm of Multi-Agent Reinforcement Learning. We will evaluate if a multi-SDN controller architecture can help provide additional improvement in our approach. 

\medbreak
\noindent {\bf Retraining:}
In this paper, we mainly described the RL-based critical flow selection policy training process as an offline task. In other words, once training is done, CFR-RL remains unmodified after being deployed in the network. However, CFR-RL can naturally accommodate future unseen traffic matrices by periodically updating the selection policy. This self-learning technique will enable CFR-RL to further adapt itself to the dynamic conditions in the network after being deployed in real networks. CFR-RL can be retrained by including new traffic matrices. For example, the outlier traffic matrices (e.g., the 235th-240th traffic matrices in Day 2) presented in Fig. \ref{exp:abilene2tm} should be included for retraining, while the generalization results shown in Section \ref{sec:generalization} suggest that retraining frequently might not be necessary. Techniques to determine when to retrain and which new/old traffic matrix should be included/excluded in/from the training dataset should be further investigated. 

The above examples are some key issues that are left for future work.

\section*{Acknowledgments}
The authors would like to thank the editors and reviewers for providing many valuable comments and suggestions.


%




\ifCLASSOPTIONcaptionsoff
  \newpage
\fi

\bibliographystyle{IEEEtran}
\bibliography{IEEEfull}
%







\begin{IEEEbiography}[{\includegraphics[width=1in,height=1.25in,clip,keepaspectratio]{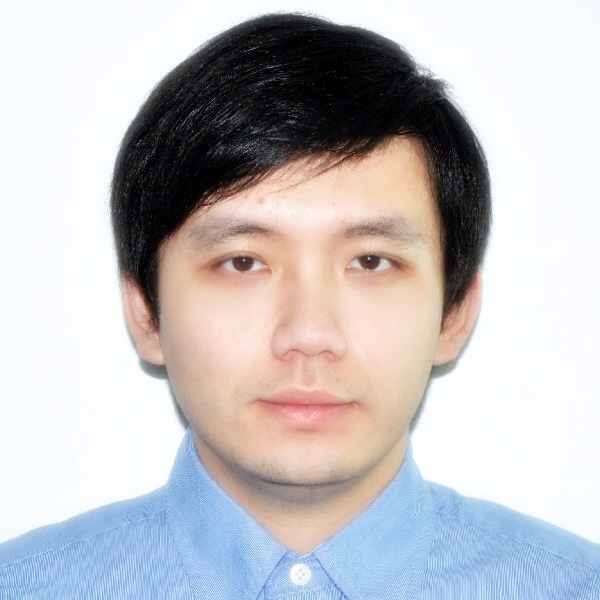}}]{Junjie Zhang} received the B.S. degree in computer science from Nanjing University of Posts \& Telecommunications, China, in 2006, the M.S. degree in computer science and the Ph.D. degree in electrical engineering from New York University, New York, NY, USA, in 2010 and 2015, respectively. 

He has been with Fortinet, Inc., Sunnyvale, CA, USA, since 2015. He holds two US patents in the area of computer networking. His research interests include network optimization, traffic engineering, machine learning, and network security. 
\end{IEEEbiography}

\begin{IEEEbiography}[{\includegraphics[width=1in,height=1.25in,clip,keepaspectratio]{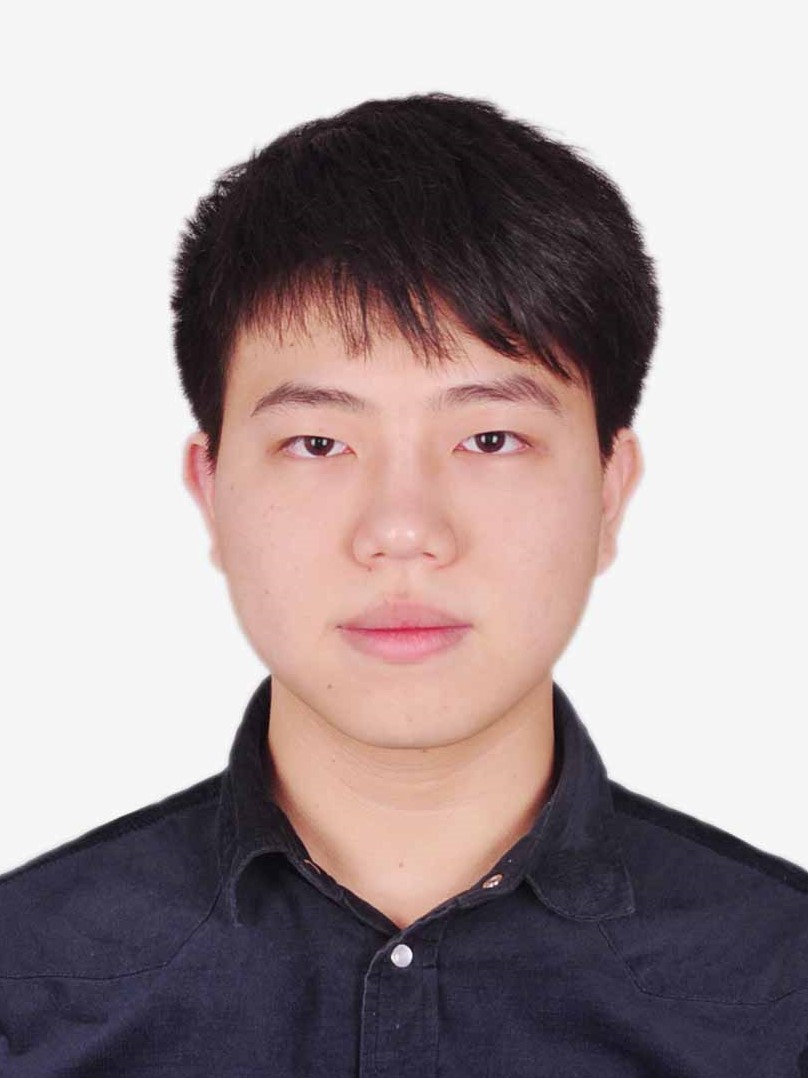}}]{Minghao Ye}received the first B.E. degree in microelectronic science and engineering from Sun Yat-sen University, Guangzhou, China, and the second B.E. degree (Hons.) in electronic engineering from Hong Kong Polytechnic University, Hong Kong, in 2017, the M.S. degree in electrical engineering from New York University, New York, NY, USA, in 2019, where he is currently pursuing the Ph.D. degree with the Department of Electrical and Computer Engineering. His research interests include traffic engineering, software-defined networks, mobile edge computing, and reinforcement learning.
\end{IEEEbiography}

\begin{IEEEbiography}[{\includegraphics[width=1in,height=1.25in,clip,keepaspectratio]{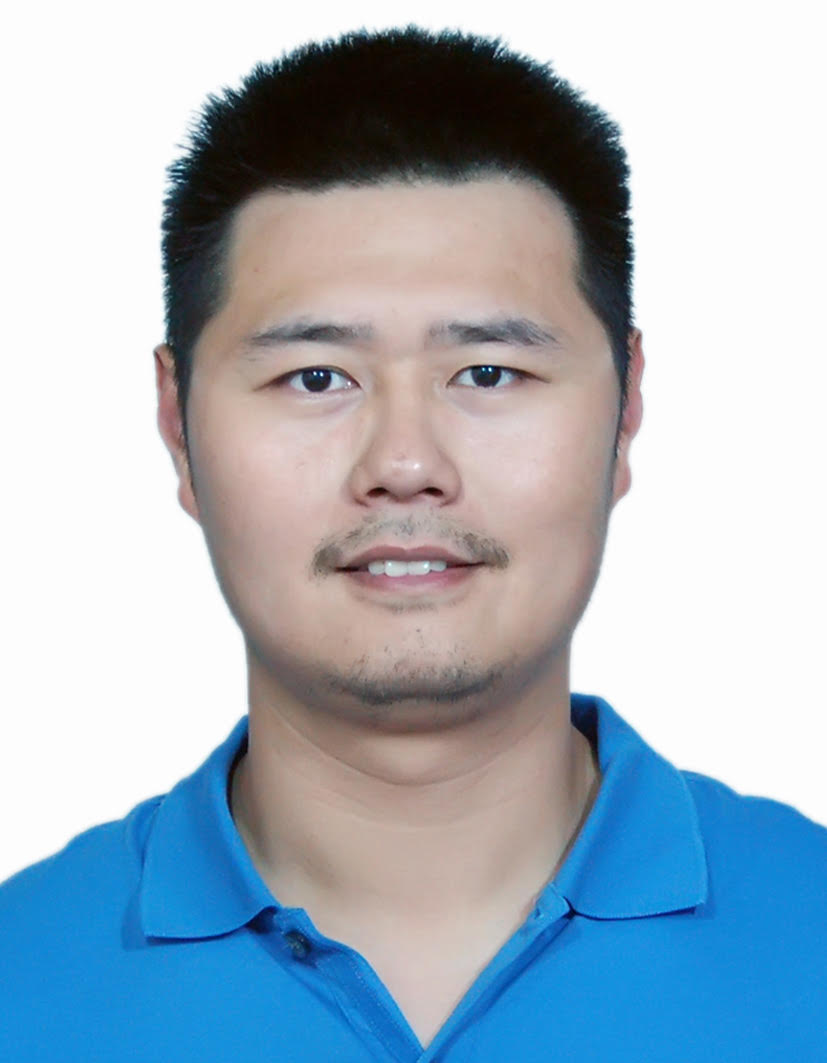}}]{Zehua Guo} (M'19-SM'20)  received a B.S. degree from Northwestern Polytechnical University, an M.S. degree from Xidian University, and a Ph.D. degree from Northwestern Polytechnical University. He is an Associate Professor at Beijing Institute of Technology. He was a Research Fellow at Department of Electrical and Computer Engineering, New York University Tandon School of Engineering, a Research Manager at ChinaCache, a Senior Software Engineer at DidiChuxing, a Post-Doctoral Research Associate at Department of Computer Science and Engineering, University of Minnesota Twin Cities, and a Visiting Associate Professor at Singapore University of Technology and Design. His research interests include software-defined networking, network function virtualization, data center network, cloud computing, content delivery network, network security, machine learning, and Internet exchange. Dr. Guo is an Associate Editor for IEEE ACCESS and the EURASIP Journal on Wireless Communications and Networking (Springer), and an Editor for the KSII Transactions on Internet and Information Systems. He was the Session Chair for the IEEE International Conference on Communications 2018 and the Technical Program Committee Member of Computer Communications (Elsevier). He is a Senior Member of IEEE.
\end{IEEEbiography}

\begin{IEEEbiography}[{\includegraphics[width=1in,height=1.25in,clip,keepaspectratio]{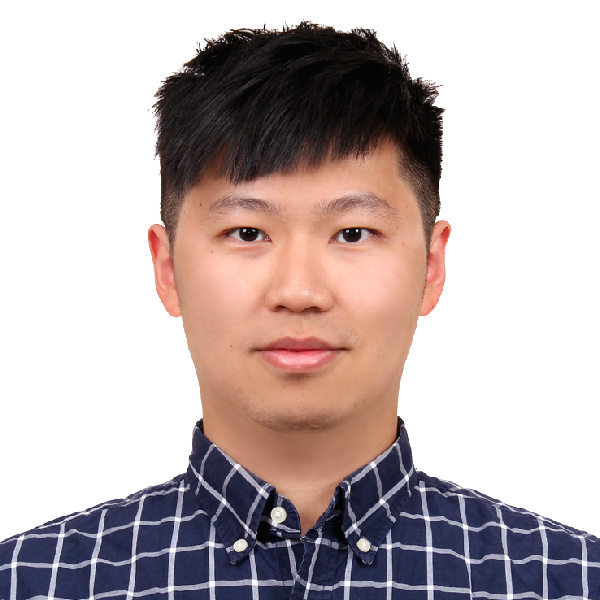}}]{Chen-Yu Yen}received the B.S. degree in electrical engineering from National Taiwan University, Taipei, Taiwan, in 2014, and the M.S. degree in electrical engineering from Columbia University in 2018. He is currently pursuing the Ph.D. degree with the Department of Electrical and Computer Engineering, New York University, New York, NY, USA. His research interests include reinforcement learning, congestion control, and practical machine learning for networking.
\end{IEEEbiography}

\begin{IEEEbiography}[{\includegraphics[width=1in,height=1.25in,clip,keepaspectratio]{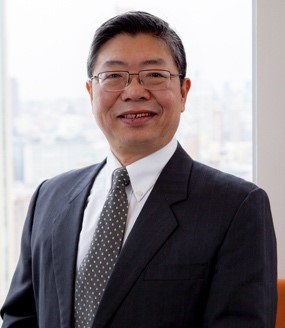}}]{H. Jonathan Chao} (M'83-F'01) received the B.S. and M.S. degrees in electrical engineering from National Chiao Tung University, Taiwan, in 1977 and 1980, respectively, and the Ph.D. degree in electrical engineering from The Ohio State University, Columbus, OH, USA, in 1985. He was the Head of the Electrical and Computer Engineering (ECE) Department at New York University (NYU) from 2004 to 2014. He has been doing research in the areas of software-defined networking, network function virtualization, datacenter networks, high-speed packet processing/switching/routing, network security, quality-of-service control, network on chip, and machine learning for networking. During 2000-2001, he was the Co-Founder and a CTO of Coree Networks, Tinton Falls, NJ, USA. From 1985 to 1992, he was a Member of Technical Staff at Bellcore, Piscataway, NJ, USA, where he was involved in transport and switching system architecture designs and application-specified integrated circuit implementations, such as the world's first SONET-like framer chip, ATM layer chip, sequencer chip (the first chip handling packet scheduling), and ATM switch chip. He is currently a Professor of ECE at NYU, New York City, NY, USA. He is also the Director of the High-Speed Networking Lab. He has co-authored three networking books, \textit{Broadband Packet Switching Technologies-A Practical Guide to ATM Switches and IP Routers} (New York: Wiley, 2001), \textit{Quality of Service Control in High-Speed Networks} (New York: Wiley, 2001), and \textit{High-Performance Switches and Routers} (New York: Wiley, 2007). He holds 63 patents and has published more than 260 journal and conference papers. He is a fellow of the IEEE and the National Academy of Inventors. He was a recipient of the Bellcore Excellence Award in 1987. He was a co-recipient of the 2001 Best Paper Award from the IEEE TRANSACTION ON CIRCUITS AND SYSTEMS FOR VIDEO TECHNOLOGY.
\end{IEEEbiography}

\end{document}